\documentclass[10pt,twocolumn,letterpaper]{article}

\usepackage{wacv}              

\usepackage{amsmath}
\usepackage{amssymb}
\usepackage{booktabs}

\usepackage[pagebackref,breaklinks,colorlinks]{hyperref}

\usepackage[capitalize]{cleveref}
\crefname{section}{Sec.}{Secs.}
\Crefname{section}{Section}{Sections}
\Crefname{table}{Table}{Tables}
\crefname{table}{Tab.}{Tabs.}

\usepackage[utf8]{inputenc} 
\usepackage[T1]{fontenc}    
\usepackage{hyperref}       
\usepackage{url}            
\usepackage{booktabs}       
\usepackage{amsfonts}       
\usepackage{nicefrac}       
\usepackage{microtype}      
\usepackage{xcolor}         
\usepackage{enumerate}
\usepackage{enumitem}

\usepackage[pdftex]{graphicx}
\graphicspath{{../pdf/}{../jpeg/}}
\DeclareGraphicsExtensions{.pdf,.jpeg,.png}

\newcommand{\Rmnum}[1]{\expandafter\@slowromancap\romannumeral #1@}
\usepackage{booktabs}
\usepackage{multirow}
\usepackage{algorithmic}
\usepackage{algorithm}
\usepackage{array}
\usepackage{subfloat}
\usepackage{textcomp}
\usepackage{stfloats}
\usepackage{verbatim}
\usepackage{wrapfig}
\usepackage{caption}
\usepackage{tabu}

\makeatletter

\def\wacvPaperID{1235} 
\def\confName{WACV}
\def\confYear{2024}

\begin{document}

\title{Med-DANet V2: A Flexible Dynamic Architecture for Efficient Medical Volumetric Segmentation}

\author{
Haoran Shen\textsuperscript{1,\thanks{Equal Contribution.\textsuperscript{\dag}Corresponding author.}}
\quad Yifu Zhang\textsuperscript{1,*}
\quad Wenxuan Wang\textsuperscript{1,*}
\quad Chen Chen\textsuperscript{2}\\
\quad Jing Liu\textsuperscript{3} 
\quad Shanshan Song\textsuperscript{1}
\quad Jiangyun Li\textsuperscript{1\dag}\\
{\textsuperscript{1}School of Automation and Electrical Engineering, University of Science and Technology Beijing} \\
{\textsuperscript{2}Center for Research in Computer Vision, University of Central Florida} \\
{\textsuperscript{3}National Lab of Pattern Recognition, Institute of Automation, Chinese Academy of Sciences} \\
\small{\texttt{m202220738@xs.ustb.edu.cn, chen.chen@crcv.ucf.edu, leejy@ustb.edu.cn}}
}

\maketitle

\begin{abstract}
Recent works have shown that the computational efficiency of 3D medical image (e.g. CT and MRI) segmentation can be impressively improved by dynamic inference based on slice-wise complexity. As a pioneering work, a dynamic architecture network for medical volumetric segmentation (i.e. Med-DANet \cite{meddanet}) has achieved a favorable accuracy and efficiency trade-off by dynamically selecting a suitable 2D candidate model from the pre-defined model bank for different slices. However, the issues of incomplete data analysis, high training costs, and the two-stage pipeline in Med-DANet require further improvement. To this end, this paper further explores a unified formulation of the dynamic inference framework from the perspective of both the data itself and the model structure. For each slice of the input volume, our proposed method dynamically selects an important foreground region for segmentation based on the policy generated by our Decision Network and Crop Position Network. Besides, we propose to insert a stage-wise quantization selector to the employed segmentation model (e.g. U-Net) for dynamic architecture adapting. Extensive experiments on BraTS 2019 and 2020 show that our method achieves comparable or better performance than previous state-of-the-art methods with much less model complexity. Compared with previous methods Med-DANet and TransBTS with dynamic and static architecture respectively, our framework improves the model efficiency by up to nearly 4.1 and 17.3 times with comparable segmentation results on BraTS 2019. 
\end{abstract}

\section{Introduction}
\label{intro}
As one of the most prevalent diseases, cancer results in numerous fatalities annually. The precise measurements of medical images is vital for accurate diagnosis and appropriate therapy planning. Traditionally, these image analysis approaches rely heavily on the doctors' clinical experience. However, it is labor-intensive and time-consuming, since a 3D volume produced by Magnetic Resonance Imaging (MRI) \cite{huo2017robust} or Computerized Tomography (CT) \cite{heller2021state} typically contains hundreds of 2D slices. Therefore, to improve the accuracy and efficiency of clinical diagnosis, automated and accurate segmentation of tumors and organs' sub-regions is a fundamental requirement for medical image analysis.

Thanks to the rapid development of deep neural networks, they have been extensively applied in the field of medical image segmentation. The mainstream segmentation methods of medical images comprise of two categories: (1) applying 2D networks for slice-by-slice predictions and (2) utilizing 3D models to process image volumes consisting of multiple slices. 2D U-Net \cite{unet}and its variants such as \cite{unet++,dcunet} are the representatives of the former category, while 3D networks like 3D U-Net \cite{3dunet} and V-Net \cite{vnet} can achieve better results owing to the associations modeling capability between different slices. Besides, as a transition from 2D methods to 3D methods, some 2.5D approaches \cite{yan2022after,hung2022cat,li2022satr} have combined information from neighboring slices to achieve better segmentation results of the current slice. Benefiting from the ability to capture long-range dependencies, many Transformer-based networks \cite{attentionunet,swinunet,unetr,swin_unetr} have begun to spring up. Nevertheless, due to the growing model scale and effective U-shaped encoder-decoder design, the high computational cost is difficult to bear in practice, especially for those 3D models.

\begin{figure*}[!h]
	\includegraphics[width=0.96\linewidth]{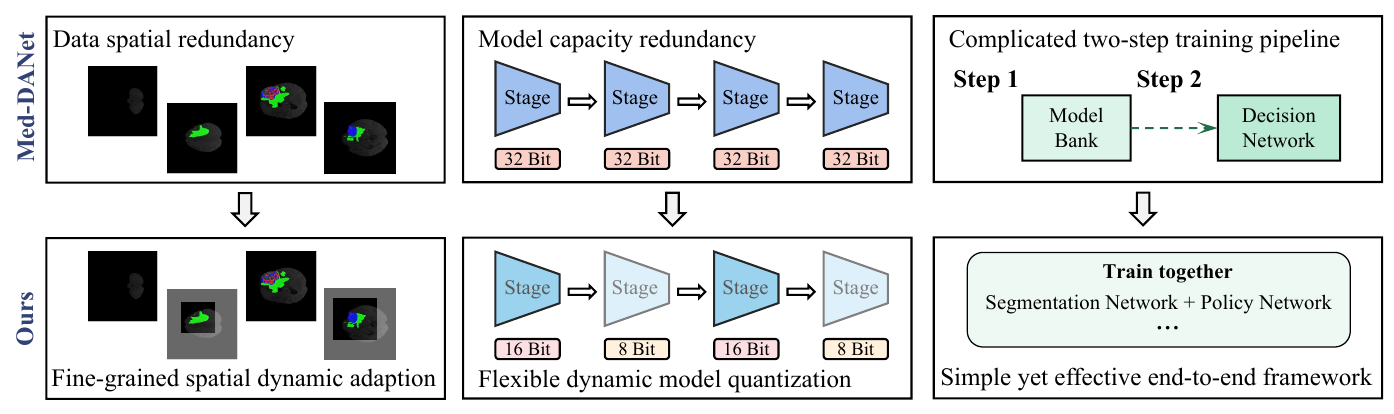}
 \vspace{-12pt}
	\caption{The comparison between the previous dynamic network Med-DANet and our proposed Med-DANet V2 (Ours).}
	\label{fig_intro}
    \vspace{-10pt}
\end{figure*}

Therefore, lightweight medical image segmentation models \cite{chen20193d,li2020memory,luo2020hdc,chen2019s3d} have become one of the research hot spots. Generally, those lightweight methods only focus on efficient structural design, ignoring the unique slice heterogeneity property of the medical images. As a pioneering work of applying dynamic inference to achieve efficient medical volumetric segmentation, the recently proposed dynamic architecture network namely Med-DANet \cite{meddanet} reveals its potential for a promising trade-off between accuracy and efficiency. Specifically, Med-DANet employs a decision network to distinguish the specific segmentation difficulty of the current input slice and accordingly selects a suitable candidate network from the pre-defined model bank for inference. In particular, the training of the decision network is supervised by a comprehensive choice metric, which is computed according to the accuracy and complexity of the incorporated candidate models. Nevertheless, as illustrated in Fig. \ref{fig_intro}, Med-DANet mainly has drawbacks in terms of three aspects: (1) Med-DANet lacks comprehensive dynamic and efficient design considering the medical data properties. Although Med-DANet chooses to directly skip to generate the segmentation mask that is filled with only background category, it ignores the spatial redundancy in those simple images (i.e., reaching precise foreground segmentation is feasible utilizing only a portion of the image area); (2) Med-DANet needs to train multiple models as candidate networks in the pre-defined model bank, which results in high training cost and limited scalability, and it also does not consider further lightweight design of each candidate network. (3) As the decision network requires a pre-calculated comprehensive choice metric for supervision, Med-DANet must be trained with a complicated two-stage pipeline, which is unfriendly for practical deployment. 

In this paper, we further explore the potential of dynamic inference in medical volumetric segmentation tasks. 
To the best of our knowledge, we are the first to unify the spatial-wise dynamic adaption and dynamic model quantization to handle the slice heterogeneity problem in MRI/CT data. 
Compared to Med-DANet, a more efficient end-to-end framework named Med-DANet V2 is proposed to get closer to clinical applications. 
To achieve more fine-grained spatial-wise dynamic adaption, we minimize input image redundancy through dynamic resolution selection and crop position determination, thereby reducing the computational complexity of the entire architecture. 
For more flexible dynamic architecture adaption, a stage-wise bit selection module is inserted into the segmentation model for dense model quantization, pursuing an extremely efficient model structure. 
Besides, our Med-DANet V2 has strong compatibility and scalability. 
The basic segmentation model can be replaced with any 2D network for various accuracy and efficiency requirements.

In summary, the main contributions are as follows:

\setlist{nolistsep}
\begin{itemize}[noitemsep,leftmargin=*]

	\item Aiming at the slice heterogeneity in medical volumes, this paper explores the potential of dynamic inference from the perspective of data properties and model structure, resulting in a promising trade-off between volumetric segmentation accuracy and efficiency.

	\item By introducing the proposed decision network and bit-width selector, we unify the spatial dynamic adaption and dynamic model quantization to realize a highly efficient medical image segmentation task.

	\item Our proposed Med-DANet V2 exhibits strong compatibility and scalability. The framework can be compatible with any 2D segmentation network to meet various accuracy and efficiency requirements. 

    \item Extensive experiments on the MRI benchmark datasets (BraTS 2019 and BraTS 2020 for brain tumor segmentation) demonstrate that our method reaches competitive or better performance than previous state-of-the-art methods with much less model complexity.
	
\end{itemize}

\section{Related Work}
\label{relatedwork}

\noindent \textbf{Static and Lightweight Methods for Medical Image Segmentation.} Recently, U-Net \cite{unet} and its variants \cite{unet++,dcunet,transbts,swinunet} have shown promising performance in medical image segmentation task. However, the high computational complexities of these models pose a great challenge for timely clinical diagnosis. Consequently, great efforts have been dedicated to designing lightweight networks for more efficient segmentation. For instance, S3D-UNet \cite{chen2019s3d} leverages separable 3D convolution to improve model efficiency. DMFNet \cite{chen20193d} takes advantage of a 3D dilated multi-fiber network to achieve the trade-off between model efficiency and accuracy in 3D MRI brain tumor segmentation tasks. HDCNet \cite{luo2020hdc} replaces 3D convolutions with a novel hierarchical decoupled convolution module to achieve a lightweight yet efficient pseudo-3D model.

\noindent \textbf{Spatial-wise Dynamic Networks.}
Most spatial-wise dynamic methods are designed for classification tasks \cite{adafocusv2,liang2022not,chen2021dynamic,xie2020spatially}, and only a few works pursue to implement the dynamic inspiration in fine-grained segmentation tasks. For example, Huang et al. \cite{huang2019uncertainty} employ a two-stage method. First, the input image is processed by a fast and small-scale model, then the uncertain regions are re-evaluated by a large network. Li et al. \cite{li2019multi} introduce data-dependent routes to adapt to the scale distribution of each image. Verelst et al. \cite{verelst2022segblocks} propose a dynamic block-based framework, where image blocks are downsampled based on their complexity. 
Although these spatial-wise dynamic methods on natural images adopt diverse strategies for different regions, the entire input is essentially segmented. However, the characteristics of having a large area of background regions (even pure background) in 3D medical data determine that there is great potential in segmenting mainly the foreground regions, which remains unexplored by previous researchers.

\noindent \textbf{Network Quantization.}
As an effective method for model compression, quantization reduces the model size and memory requirements, thereby accelerating inference. Although the quantization technique in the natural image field has attracted widespread attention in the community, its potential in medical image analysis has rarely been explored. Additionally, the utilization of low bit-width methods for network quantization leads to significant performance degradation \cite{choi2018pact,askarihemmat2019u,zhang2021medq,hubara2016binarized}. 
Thus, some mixed-precision bit-width methods \cite{guo2022mixed,chen2021towards,wu2018mixed,lou2019autoq} have become popular to improve quantization for a better trade-off. Although applying this method to medical image segmentation models can reduce the varying redundancy of different layers, it cannot adaptively allocate appropriate quantization bit widths for specific medical slices. Therefore, dynamic quantization in the medical field remains challenging and to be exploited, especially for medical image segmentation tasks that require separating tissues and organs with high precision.

\section{Methodology}
\label{methodology}
\subsection{Preliminary: Designing Details of Med-DANet}
Med-DANet \cite{meddanet} is a dynamic architecture network that aims to handle 3D medical images where segmentation targets are sparsely distributed among slices. Given a 3D volume, a slice-specific decision is learned by the Decision Network to dynamically select a suitable model from the pre-defined model bank for subsequent segmentation. The Decision Network undertakes a $n+1$-class classification task, where the $n+1$ categories represent $n$ candidate networks and a skip process. The Decision Network and model bank are respectively formulated as follows
\begin{align}
    \label{eq1}\mathcal{D}(x)&=\{\hat{D}|x;\theta\},\\
    \label{eq2}\mathcal{B}(x)&=[\varnothing,M_1(x),M_2(x),...,M_n(x)].
\end{align}

where $\theta$ denotes the parameters of Decision Network and $\hat{D}$ is the prediction of $\mathcal{D}(x)$. $M_{1} \sim M_{n}$ indicate the candidate models and $(\varnothing)$ represents the skip operation.

During the training process, Med-DANet proposes a comprehensive choice metric to supervise the framework, which is computed by the accuracy and complexity of the incorporated models. The metric is calculated as follows.
\begin{equation}
\footnotesize
    \label{eq4}
    D=\left\{
    \begin{array}{lr}
        0, & P_f < 1 \\
        argmax((1-\alpha)*S_i + \alpha*softmax(\frac{1}{F_i})) + 1, & P_f \geqslant 1
    \end{array},
    \right.
\end{equation}
where $S_i$ and $F_i$ are respectively the Dice Score and FLOPs of each candidate model $M_i$. $P_f$ denotes the number of foreground target pixels.

\noindent \textbf{Limitations of Med-DANet.}
(1) Although Med-DANet selects models of corresponding scales based on the difficulty of slices, it ignores handling simple images with various background regions from the input perspective.
(2) The model bank with multiple models must be pre-established, which requires a high training cost.
(3) The pre-calculated comprehensive choice metric requires Med-DANet to be trained using a complex two-stage pipeline, which may pose challenges for users.

Given these limitations, a more flexible and unified dynamic one-stage model architecture named Med-DANet V2 is developed. In terms of the input data, more fine-grained spatial-wise dynamic adaption will be determined by two decision modules. As for the model capacity, the bit-width quantization will be considered to achieve more flexible dynamic architecture adaption.

\begin{figure*}[htbp]
	\includegraphics[width=1\linewidth]{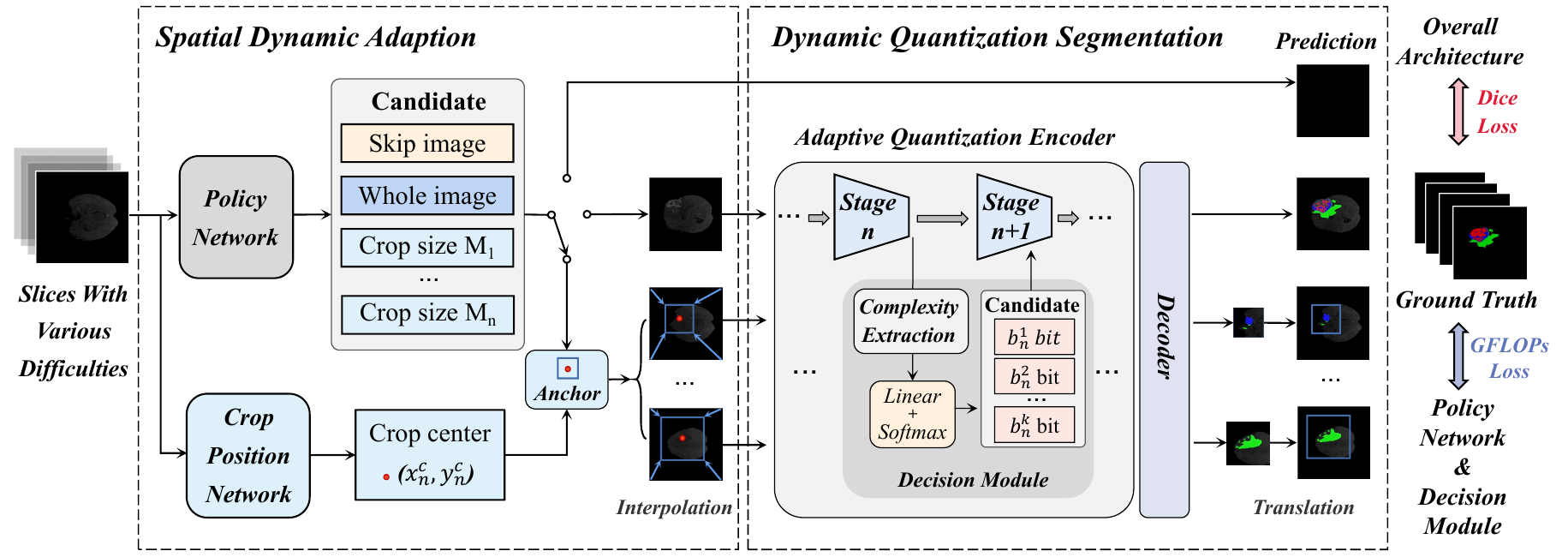}
 \vspace{-0.7cm}
	\caption{The illustration of the our Med-DANet V2. The Policy Network takes a 2D image slice as input and generates a choice depending on the segmentation difficulty of the current slice. Based on the optimal choice made by the Policy Network, our method can adaptively decide whether to skip the current slice (i.e. directly generating the result with all zero -- "background" class) or send the input with a suitable resolution to the Dynamic Quantization Model (with the cropping region determined by the Crop Position Network). A stage-wise Decision Module is inserted into the Dynamic Quantization Model for suitable model capacity selection. By unifying data-architecture dynamic inference, our method can achieve accurate and efficient segmentation. }
	\label{fig_methodology}
    \vspace{-10pt}
\end{figure*}

\subsection{Overall Architecture}
The overall architecture of the proposed Med-DANet V2 is depicted in Fig. \ref{fig_methodology}. In general, our framework consists of the data decision part (i.e. Policy Network ($P$) and Crop Position Network ($C$)) and the Dynamic Quantization Network ($Q$). To deal with the medical datasets where segmentation targets are sparsely distributed among inter and intra-slices, a slice-specific decision dynamically selects a suitable crop size (i.e. obtained by $P$) and a crop-center (i.e. obtained by $C$) to locate the cropping image. Then the cropped image will be segmented by the Dynamic Quantization Network, which can automatically allocate the bit-width inside the model to achieve different model scales. Dynamic choice of architecture and the subsequent segmentation task are formulated by Eq. \ref{metric1}:
\begin{equation}\label{metric1}
		y = Q[P({x}) \ o\ C({x})] \\
\end{equation}  
where $x$ denotes the input image and $y$ is the corresponding prediction. $Q[ P \circ C]$ indicates feeding the slice of a dynamic crop into the dynamic quantization model. In the rest of this section, the designing details about Spatial Dynamic Adaption (Sec. \ref{sec:dynamicresolution}), Dynamic quantization (Sec. \ref{sec:dynamicquantization}), and Training/Inference Strategy (Sec. \ref{sec:trainingandinference}) will be explained.
As a whole, the Policy Network combined with the Crop Position Network will comprehensively consider the foreground and background regional distribution inter and intra slice, which can make the most appropriate crop choice(i.e. containing skip). Regarding the Dynamic Quantization Model, any 2D network can be tailored as the foundation to flexibly accommodate diverse accuracy and efficiency demands. More discussions on the model choices and ablation study are presented in Sec. \ref{sec:AblationStudies}.

\subsection{Fine-grained Spatial-wise Dynamic Adaption}
\label{sec:dynamicresolution}
Due to the large amount of redundant background regions in some simple medical image slices, feeding only the foreground part into the network is sufficient for obtaining good results. Besides, minimizing the proportion of the meaningless regions can make the segmentation model more focused on significant feature extraction (e.g. features of crucial tumor regions), accordingly achieving better performance. To achieve this, two decision networks are introduced to determine the resolution size and center point of the cropped image tile for the following segmentation. Then the selected slice region based on the decisions is sent to the segmentation network to obtain the final result.

Following Med-DANet, the high-efficiency ShuffleNet V2 \cite{shufflenetv2} is selected as our Policy Network and Crop Position Network. Hence, only negligible computation cost is added to the entire framework. The Policy Network acts as a classifier to distinguish  $n+1$ categories, which refer to $n$ candidate resolutions and a direct skip operation. Besides, we use Gumbel-Softmax Sampling \cite{gumbelsoftmax} to enable the network optimization process by standard back-propagation. Specifically, when choosing a skip operation, the whole architecture will skip the subsequent segmentation process and produce the segmentation maps with the background class. If the choice of full-resolution image is made, the input slice will be straightly fed into the Dynamic Quantization Model for final prediction. Otherwise, the Crop Position Network comes into effect and undertakes a regression task to determine the center point for cropping, which is used before and after the segmentation model for resolution downsampling and restoration.

For the process of altering resolution, we follow AdaFocus V2 \cite{adafocusv2} to use differentiable bilinear interpolation for downsampling. Assuming the sizes of the whole input and the cropped image are $H \times W$ and $P_n \times P_n$. Each pixel $(x_n,y_n)$ in the cropped patch can be expressed by the center position $(x^c_n,y^c_n)$ and a fixed offset $o_{ij}$:
\begin{equation}
\footnotesize
\label{metric2}
		(x_n,y_n) = (x^c_n,y^c_n) + o_{ij},\\
		o_{ij} \in {\{ -\frac{P_n}{2}, -\frac{P_n}{2} + 1, \ldots,\frac{P_n}{2} \}}^2.
\end{equation}
Then the actual values of the cropped pixels are calculated by the continuous $(x^c_n,y^c_n)$ via interpolation. To effectively adapt the dynamic resolution cropping strategy to segmentation tasks, we propose a reverse process with little modification in the segmentation network for the generation of full-resolution segmentation masks. Concretely, feature maps of the cropped input image are initially positioned at the upper left corner of a full-sized map with all elements set to zero. Afterward, an affine translation is performed on the full-sized map according to the center point position. Intuitively, directly interpolating the cropped segmentation result to full resolution may not provide precise results. Instead of designing a brand-new module to refine, we interpolate the feature maps before the last decoder stage in the Dynamic Quantization Model, and then use size-independent last decoder stage to obtain the final segmentation results.

\subsection{Flexible Dynamic Architecture Adaption}
\label{sec:dynamicquantization}
Previous quantization works \cite{askarihemmat2019u,rastegari2016xnor,choi2018pact} have normally quantized the network with a fixed bit-width. Nevertheless, different network stages exhibit different degrees of quantization sensitivity (performance drop from a fixed-bit quantization). Therefore, we aim to dynamically assign bit widths for the layers at different stages based on the feature complexity of the slices, avoiding the significant accuracy reduction and computing resources waste of uniformly quantizing a fixed bit-width network. To achieve high average flops-reduction with less accuracy loss, we propose a novel stage-distribution dynamic quantization method that allocates optimal bits to quantize each operator at different stages of the model.

Generally, to replace the majority of floating-point operations with lower-bit
operations in a universal model, the input feature and weight of each layer can be respectively quantized \cite{choi2018pact,jung2019learning}. Given the operator weight of the $j$-th layer, it will be quantized to $Q_{b}\left(\boldsymbol{w}^{j}\right)$:
\begin{equation}
Q_{b}\left(\boldsymbol{w}^{j}\right)=\left\lfloor\operatorname{clip}\left(\boldsymbol{w}^{j}, a\right) \cdot \frac{r(b)}{a}\right\rceil \cdot \frac{a}{r(b)}.
\end{equation}

$w^{j}$ is first truncated with the clip function and scaled to [-1, 1] with the scale parameter $a$. Then, $w^{j}$ is scaled to the integer range $[-r(b), r(b)]$ of the given bit-width, where $r(b)=2^{b-1}$. We use the simplest quantization strategy, the weight scale parameter $a$ is determined simply by $a = max w^{j}$.

In our framework, each convolution and linear layer will be quantized, one of which is in turn selected by a bit selector consisting of K bit-width quantization candidates, as illustrated in Fig. \ref{fig_methodology}. To realize the bit-width selection in the model's different stages, we use a lightweight bit selection module that assigns quantization bits adaptively according to the feature complexity extracted from the slice (i.e. gradient of spatial feature and standard deviation of channel feature). The bit-width selector is applied before each stage and the highest probability bit-width will be selected.

The selector outputs a quantization function of the max probability, which is a discrete non-differentiable process and cannot be optimized end-to-end. The straight-through estimator trick \cite{bengio2013estimating} substitutes the discrete process to make the process differentiable, as shown in Eq. \ref{metric4}, during forward-propagation and back-propagation:
\begin{equation}
\label{metric4}
    b_{n}^{k^{*}}=\left\{
    \begin{array}{lr}
        argmax_{b_{n}^{k}} P_{b_{n}^{k}}\left(\boldsymbol{x}^{n}\right) & forward \\
        \sum_{k=1}^{K} b_{n}^{k} \cdot P_{b_{n}^{k}}\left(\boldsymbol{x}^{n}\right) & backward
\end{array}\right.
\end{equation}
where $P_{b_{n}^{k}}$ is the probability assigned to the bit-width ${b_{n}^{k}}$ of the $n$-th stage among K numbers of candidate.

Previous works \cite{mishra2017apprentice,romero2014fitnets} have mainly focused on optimizing a fixed bit-width quantized network in a static manner, by taking the original network as the teacher model with a knowledge distillation loss and a conventional pixel-wise supervision loss. 
To realize a more efficient quantization strategy to adaptively adjust the model capacity in a dynamic way, we directly employ the obtained computational complexity of the dynamic quantized model as the loss value. 
To achieve a better trade-off between the computational cost and restoration performance, the bit-widths of quantization modules with a larger impact on precision should be allocated a large number of quantization bits, resulting in the following GFLOPs loss:
\begin{equation}
        GFLOPs\ Loss=\sum_{0}^{N}GFLOPs(Q_{b}({w}^{j}) ,b_{j}^{k^{*}}).
\end{equation}
where $N$ represents the stage numbers.
\subsection{Training and Inference Strategy}
\label{sec:trainingandinference}

\noindent \textbf{Training.} 
 The two decision networks and the Dynamic Quantization Model are trained at three different periods. Firstly, we train the full-precision segmentation model to ensure benchmark accuracy. Next, we train the dynamic quantization model, which adaptively assigns bit-widths based on feature complexity and adjusts the weights of the entire model accordingly. Finally, we train the entire architecture, including the decision networks and quantization model together, pursuing an optimal balance between accuracy and computational complexity.
To address the problem of inter-slice image and intra-slice region distribution (i.e. background occupies a significant portion or the entirety of a slice, which is a substantial part of the dataset), the combination of Dice Loss (Eq. 9) and GFLOPs Loss (Eq. 8) is employed to meet well trade-off between segmentation accuracy and model complexity. Specifically, the Dice Loss is used to supervise the entire network, while GFLOPs Loss is utilized to supervise the selection of the Policy Network and Dynamic Quantization Model.
\begin{align}
        \label{dice}
    Dice\ Loss&= \sum_{i=1}^{C}(1-\frac{2|pred_i \cap truth_i|}{|pred_i|+|truth_i|}),\\
    Loss&=Dice\ Loss+\lambda\ GFLOPs\ Loss.
\end{align}

where $Loss$ represents the overall loss, $C$ denotes class numbers and $\lambda$ is a weight parameter. 

\noindent \textbf{Inference.} After the one-stage training phase mentioned above, the well-trained decision networks and Dynamic Quantization Model are cascaded sequentially to achieve efficient inference. Given a 2D slice as input, our lightweight Decision Network will decide to skip the current slice or generate a specific crop decision according to the segmentation difficulty of the current slice. A segmentation map with all zeros will be directly adopted for final results if the skip operation is determined by the Policy Network. Otherwise, based on the current slice's feature complexity, an appropriate combination of crop size and crop center point is determined by the Policy Network and Crop Position Network. Then, the cropped or whole image will be segmented by our Dynamic Quantization Model, which can allocate bit-width for different model stages according to the complexity of image features. In this way, a dynamic slice-dependent architecture with greatly improved efficiency is realized by our method in terms of both input information and model scale. On one hand, compared with the previously proposed lightweight static structure, our Med-DANet V2 simultaneously takes advantage of spatial-wise dynamic selection and dynamic architecture to adjust the inter-slice image and intra-slice regions instead of treating all inputs and regions equally. On the other hand, compared with the previously proposed dynamic methods that train multiple models for constructing a pre-defined model bank or utilize early exit to construct the cascaded dynamic architecture, our highly efficient Med-DANet V2 not only requires low training costs (i.e. only trains a single model) but also implements an extremely lightweight dynamic architecture network by adopting dynamic resolution adjustment and adaptive quantization simultaneously.

\section{Experimental Evaluation}
\subsection{Experimental Setup}
\noindent \textbf{Data and Evaluation Metric.}
The first 3D MRI dataset used for experiments is provided by the Brain Tumor Segmentation Challenge (BraTS) 2019 \cite{menze2014multimodal,bakas2017advancing,bakas2018identifying}. It comprises a training set with 335 cases and a validation set with 125 cases, each with four modalities (T1, T1c, T2, and FLAIR) rigidly aligned. The size of each modality is resampled to $240\times240\times155$.
All cases are labeled by four categories: background (label 0), necrotic and non-enhancing tumor (label 1), peritumoral edema (label 2), and GD-enhancing tumor (label 4).
The segmentation performance on three classes: enhancing tumor region (ET, label 4), regions of the tumor core (TC, labels 1 and 4), and the whole tumor region (WT, labels 1, 2, and 4) is evaluated by the Dice score and the Hausdorff distance (95\%) metrics, while FLOPs is used for computational complexity measurement. 
The second 3D MRI dataset used in our study is obtained from the Brain Tumor Segmentation Challenge (BraTS) 2020. The dataset consists of 369 cases for training and 125 cases for validation. Apart from the difference in sample quantity, the remaining information for these two MRI datasets is identical. Due to space limitations, we place the visualization results in the \textit{\textbf{supplementary material}}.

\noindent \textbf{Implementation Details.}  Our Med-DANet V2 is implemented based on PyTorch \cite{paszke2019pytorch} and trained with 4 Geforce RTX 3090 GPUs. The candidates for Policy Network are set to skip, whole image, and crop size 96, while the quantization bit selection space is set to 8 bits and 16 bits.
We trained our method for 350 epochs with a batch size of 64. The training process consists of three phases: the initial 200 epochs when the Policy Network and full precision segmentation module are trained, then the middle 100 epochs to train the employed Crop Position Network and Dynamic Quantization Model, and the final 50 epochs to train the entire framework. 
The initial learning rate is set to $2e^{-4}$ and $1e^{-4}$ for BraTS 2019 and 2020 respectively.
The Adam optimizer and poly learning rate strategy with warm-up are employed for model training. We follow the data augmentation techniques and model regularization in \cite{meddanet}. A combination of the softmax Dice Loss and GFLOPs Loss is employed, while the weight factor $\lambda$ for balancing the two losses is set to 0.06.

\noindent \textbf{Baseline Selection.} As the representation of CNNs, U-Net \cite{unet} achieves state-of-the-art performance in medical segmentation owing to its powerful information extraction ability.
Compared to the original U-Net, the modified version makes an improvement in both accuracy and complexity. Thus we choose the modified 2D U-Net with 16 base channels as the baseline in our framework. 

\subsection{Results and Analysis}

\textbf{BraTS 2019.} 
We conduct experiments on the BraTS 2019 validation set and compare our Med-DANet V2 with previous state-of-the-art (SOTA) approaches. The quantitative results are as reported in Table \ref{tab:comparison2019}. Obviously, our method achieves comparable or higher performance than previous SOTA methods with significantly less computational complexity. Specifically, our method reaches the Dice scores of $80.08\%$, $90.27\%$, $81.28\%$ on ET, WT, TC, respectively. In addition, Hausdorff Distance metrics of $3.494\%$, $5.871\%$, $6.170\%$ on ET, WT, TC prove the credibility of our segmentation results from another perspective. Besides, Med-DANet V2 has impressively lower complexity compared to other SOTA methods. For instance, Med-DANet \cite{meddanet} has a computational complexity \textbf{4.1} times of ours per slice, while the model complexity of TransBTS \cite{transbts} is incredibly \textbf{17.3} times of ours. These findings strongly support the effectiveness of unifying data and model structure for dynamic inference.

\begin{table*}[!t]
  \caption{Performance Comparison on BraTS 2019 Validation Set.}
  \label{tab:comparison2019}
  \centering
  \small
  \setlength\tabcolsep{5.4pt}%
  \begin{tabular}{ccccccccc}
    \toprule
    \multirow{2}{*}{Backbone} & \multicolumn{3}{c}{Dice Score (\%) $\uparrow$} & \multicolumn{3}{c}{Hausdorff Dist. (mm) $\downarrow$} & \multicolumn{2}{c}{FLOPs (G) $\downarrow$} \\
    \cmidrule(r){2-4} \cmidrule(r){5-7} \cmidrule(r){8-9}
             &  ET &  WT &  TC &  ET &  WT & TC & Per Case & Per Slice \\
    \midrule
        3D U-Net \cite{3dunet}  & 70.86 & 87.38 & 72.48 & 5.062 & 9.432 & 8.719 & 1,669.53 & 13.04 \\
        V-Net  \cite{vnet}    & 73.89 & 88.73 & 76.56 & 6.131 & 6.256 & 8.705 & 749.29 & 5.85 \\
        Attention U-Net  \cite{attentionunet}   & 75.96 & 88.81 & 77.20 & 5.202 & 7.756 & 8.258 & 132.67 & 1.04 \\
        Wang et al. \cite{3dunet}         & 73.70 & 89.40 & 80.70 & 5.994 & 5.677 & 7.357 & - & - \\
        Chen et al.  \cite{chen2019aggregating}   & 74.16 & 90.26 & 79.25 & 4.575 & \textbf{4.378} & 7.954 & - & - \\
        Li et al. \cite{li2019multi}           & 77.10 & 88.60 & 81.30 & 6.033 & 6.232 & 7.409 & -  & - \\
        TransUNet  \cite{chen2021transunet}   & 78.17 & 89.48 & 78.91 & 4.832 & 6.667 & 7.365 & 1205.76 & 9.42 \\
        Swin-UNet  \cite{swinunet}   & 78.49 & 89.38 & 78.75 & 6.925 & 7.505 & 9.260 & 250.88 & 1.96 \\
        TransBTS  \cite{transbts}   & 78.36 &  88.89 & \textbf{81.41} & 5.908 & 7.599 & 7.584 & 333.09 & 2.60 \\
        Med-DANet  \cite{meddanet}   & 79.99 &  90.13 & 80.83 & 4.086 & 5.826 & 6.886 & 77.78 & 0.61 \\
    \midrule
     \bf{Ours}          & \textbf{80.08} & \textbf{90.27} & 81.28 & \textbf{3.494} & 5.871 & \textbf{6.170} & \textbf{19.36} & \textbf{0.15}\\    
    \bottomrule
  \end{tabular}

\end{table*}

\textbf{BraTS 2020.} 
 The comparisons between our Med-DANet V2 with previous SOTA approaches on BraTS 2020 are as reported in Table \ref{tab:comparison2020}. 
 Our method reaches $80.38\%$, $90.14\%$, $81.16\%$ on Dice scores and $10.172\%$, $6.153\%$, $8.221\%$ on Hausdorff Distance with a GFLOPs of 33.85 per case. It shows that our framework can significantly improve inference efficiency while maintaining comparable segmentation accuracy. 
 For example, Med-DANet \cite{meddanet} has a computational complexity \textbf{2.3} times of our method, while the complexity of TransBTS \cite{transbts} is incredibly \textbf{10.0} times of ours.

\begin{table*}[!t]
  \caption{Performance Comparison on BraTS 2020 Validation Set.}
  \label{tab:comparison2020}
  \centering
  \small
  \setlength\tabcolsep{5.4pt}%
  \begin{tabular}{ccccccccc}
    \toprule
    \multirow{2}{*}{Backbone} & \multicolumn{3}{c}{Dice Score (\%) $\uparrow$} & \multicolumn{3}{c}{Hausdorff Dist. (mm) $\downarrow$} & \multicolumn{2}{c}{FLOPs (G) $\downarrow$} \\

    \cmidrule(r){2-4} \cmidrule(r){5-7} \cmidrule(r){8-9}
             &  ET &  WT &  TC &  ET &  WT & TC & Per Case & Per Slice \\
    \midrule
        3D U-Net \cite{3dunet}  & 68.76 & 84.11 & 79.06 & 50.983 & 13.366 & 13.607 & 1,669.53 & 13.04 \\
        V-Net  \cite{vnet}      & 61.79 & 84.63 & 75.26 & 47.702 & 20.407 & 12.175 & 749.29 & 5.85 \\
        Deeper V-Net  \cite{vnet}  & 68.97 & 86.11 & 77.90 & 43.518 & 14.499 & 16.153 & - & -\\
        3D Residual U-Net \cite{zhang2018road}   & 71.63 & 82.46 & 76.47 & 37.422 & 12.337 & 13.105 & 407.37 & 3.18\\
        Liu et al. \cite{liu2020brain}   & 76.37 & 88.23 & 80.12 & 21.390 & 6.680 & \textbf{6.490} & - & -\\
        Ghaffari et al. \cite{ghaffari2020brain}   & 78.00 & 90.00 & 82.00 & - & - & - & - & -\\
        TransUNet  \cite{chen2021transunet}   & 78.42 & 89.46 & 78.37 & 12.851 & \textbf{5.968} & 12.840 & 1205.76 & 9.42  \\
        Swin-UNet  \cite{swinunet}   & 78.95 & 89.34 & 77.60 & 11.005 & 7.855 & 14.594 & 250.88 & 1.96 \\
        TransBTS  \cite{transbts} & 78.50 & 89.00 & \textbf{81.36} & 16.716 & 6.469 & 10.468 & 333.09 & 2.60\\
        Med-DANet  \cite{meddanet}   & \textbf{80.57} & \textbf{90.28} & 81.34 & \textbf{6.474} & 6.718 & 7.416 & 77.71 & 0.61 \\
    \midrule
     \bf{Ours}          & 80.38 & 90.14 & 81.16 & 10.172 & 6.153 & 8.221 & \textbf{33.85} & \textbf{0.26}\\    
    \bottomrule
  \end{tabular}

\end{table*}

\subsection{Ablation Study}
\label{sec:AblationStudies}

To better understand the proposed methods, we conduct ablation experiments on components and settings. All studies are conducted with a U-Net baseline on the BraTS 2019 training set under the five-fold cross-validation setting unless specified otherwise.

\noindent \textbf{Ablation Study on the Dynamic Architecture.} 
We first conduct effectiveness validation of Med-DANet V2 framework for the dynamic adaption in the spatial domain (S) and quantization architecture (Q). As shown in Table \ref{tab:ablation1}, our method achieves significant improvements in both segmentation accuracy and model efficiency compared to baseline. The fine-grained spatial-wise dynamic adaption reduces the computational complexity by nearly 2.2 times and improves all the Dice Scores. The flexible dynamic architecture adaption reduces computational complexity by about 1.9 times without sacrificing too much precision. By integrating the advantages of both aspects, the results of accuracy and efficiency prove the significance of taking spatial and architecture-wise adaption into consideration for dynamic medical inference.

\noindent \textbf{Ablation Study on the Weight Factor $\lambda$ of the GFLOPs Loss.}
In this section, we explore the optimal balance between model complexity and performance. As described in Sec. \ref{sec:trainingandinference}, we introduce $\lambda$ to control the computational complexity of the model. Ablation results are listed in Table \ref{tab:ablation2}, which indicates that $\lambda=0.06$ is the optimal weight for achieving the best trade-off. Continuing to increase the constraint on computational complexity will make the model select some slices to use higher resolution while they perform better in low resolution. And also the high-resolution slices with small foregrounds may weaken the model's ability to focus on large targets.

\noindent \textbf{Ablation Study on the Policy Space.} 
We explore the effect of policy space by using different combinations of resolution and skipping. The resolution candidates are set to $2^n$, namely 32, 64, 96, and the whole image. As shown in Table \ref{tab:ablation3}, the policy space of $\{$Skip, 96, Whole$\}$ achieves the best trade-off in accuracy and efficiency compared to all the alternatives. And the further increase of candidates in policy space may not achieve better performance.

\noindent \textbf{Ablation Study on the Quantization Space.} 
In this section, we investigate the impact of using different combinations of Quantization Space, as displayed in Table \ref{tab:ablation4}. Intermediate quantization bits such as 8, 12, and 16 are selected as candidates for the quantization space. Among them, \{8,16\} achieves the best balance between accuracy and computational complexity. Setting the quantization space as \{8,12,16\} and \{12,16\} would increase the model's capacity, but the fixed constraints of $\lambda$ introduce biases towards smaller values in both the decision space and quantization space, resulting in inadequate segmentation performance. Although \{8,12\} has the similar computational complexity as \{8,16\}, its upper limit of 12 bits hinders the segmentation for some complex slices.

\noindent \textbf{Ablation Study on Different Baselines.} 
We choose two baselines (the modified 2D UNet and the 2D version of lightweight TransBTS) to verify the compatibility and scalability of our framework because they represent two popular architectures (i.e. convolutional neural networks and vision Transformers).
As shown in Table \ref{tab:ablation5}, by unifying spatial adaption and dynamic quantization architecture, our framework significantly surpasses both baselines in computational complexity with comparable or even better performance.

\begin{table*}[t]
    \centering
    \footnotesize
    \begin{subtable}[t]{0.3\linewidth}
        \centering
        \renewcommand{\arraystretch}{1.3}
        \setlength{\tabcolsep}{1.0mm}{
        \begin{tabular}{cccccc}
        \toprule
        \multicolumn{2}{c}{Dynamic}  & \multicolumn{3}{c}{Dice Score (\%) $\uparrow$}  & FLOPs (G) $\downarrow$ \\

        \cmidrule(r){1-2} \cmidrule(r){3-5} 
           S & Q &  ET &  WT &  TC & Per Slice  \\
        \midrule
             -  & - & 78.31 & 90.61 & 82.59  & 17.07  \\
         \checkmark   &   & 78.40 & 90.95 & \textbf{83.39}  & 7.87   \\
        & \checkmark  & 77.10 & 90.76 & 81.86  & 9.01   \\
          \checkmark  & \checkmark   & \textbf{78.55} & \textbf{91.01} & 82.85  & \textbf{2.75}   \\
          
        \bottomrule
        \end{tabular}}
        \caption{Ablation Study on the Components of Med-DANet V2 Architecture.}
        \label{tab:ablation1}
    \end{subtable}
    \begin{subtable}[t]{0.3\linewidth}
        \centering
        \renewcommand{\arraystretch}{1.3}
        \setlength{\tabcolsep}{0.7mm}{
        \begin{tabular}{ccccc}
        \toprule
        \multirow{2}{*}{$\lambda$} & \multicolumn{3}{c}{Dice Score (\%) $\uparrow$}  & FLOPs (G) $\downarrow$ \\
        \cmidrule(r){2-4} 
            &   ET &  WT &  TC  & Per Slice \\
        \midrule
        0.1  & 78.5 & 90.86 & \textbf{82.88}  & \textbf{2.17}  \\
        0.08    & 78.43 & 90.93 & 82.80  & 2.23   \\
        0.06   & \textbf{78.55} & \textbf{91.01} & 82.85  & 2.75   \\
        0.04    & 78.32 & 90.43 & 82.36 & 4.12   \\
        
        \bottomrule
    \end{tabular}}
    \captionsetup{width=0.9\textwidth}
    \caption{Ablation Study on the Weight Factor $\lambda$ of the GFLOPs Loss.}
    \label{tab:ablation2}
    \end{subtable}
    \begin{subtable}[t]{0.3\linewidth}
        \centering
        \renewcommand{\arraystretch}{1.3}
        \setlength{\tabcolsep}{0.7mm}{
        \begin{tabular}{ccccc}
        \toprule
    Quantization  & \multicolumn{3}{c}{Dice Score (\%) $\uparrow$}  &FLOPs (G) $\downarrow$ \\
        \cmidrule(r){2-4} 
          Space   &  ET &  WT &  TC  & Per Slice  \\
        \midrule
        $\{8,12\}$   & 77.07 & 90.41 & 82.63 & 2.74 \\
        $\{8,16\}$   & \textbf{78.55} & \textbf{91.01}  & \textbf{82.85} & 2.75  \\
        $\{12,16\}$    & 76.81 & 90.27 & 81.58 & \textbf{2.69}   \\
        $\{8,12,16\}$  & 77.69 & 90.59 & 82.73  & 3.25 \\
        \bottomrule
    \end{tabular}}
    \caption{Ablation Study on the Quantization Space.}
    \label{tab:ablation4}
    \end{subtable}
    \begin{subtable}[t]{0.47\linewidth}
        \centering
        \renewcommand{\arraystretch}{1.39}
		\setlength{\tabcolsep}{2.2mm}{
        \begin{tabular}{ccccc}
        \toprule
        \multirow{2}{*}{Policy Space} & \multicolumn{3}{c}{Dice Score (\%) $\uparrow$}  & FLOPs (G) $\downarrow$  \\
        \cmidrule(r){2-4} 
             &  ET &  WT &  TC  & Per Slice  \\
        \midrule
        $\{$Skip, 32, Whole$\}$   & 78.24 & 89.41 & 81.37  & \textbf{2.67} \\
        $\{$Skip, 64, Whole$\}$    & \textbf{78.89} & 90.60 & 82.81  & 3.63   \\
        $\{$Skip, 96, Whole$\}$   & 78.55 & \textbf{91.01} & \textbf{82.85}  & 2.75  \\
        $\{$Skip, 64, 96, Whole$\}$  & 78.64 & 90.52 & 82.26  & 3.67 \\
        \bottomrule
  \end{tabular}}
    \caption{Ablation Study on the Policy Space.}
  \label{tab:ablation3}
    \end{subtable}
    \begin{subtable}[t]{0.47\linewidth}
        \centering
        \renewcommand{\arraystretch}{1.3}
		\setlength{\tabcolsep}{2.2mm}{
\begin{tabular}{cccccc}
    \toprule
    \multirow{2}{*}{Backbone} & \multirow{2}{*}{Method} & \multicolumn{3}{c}{Dice Score (\%) $\uparrow$}  & FLOPs (G) $\downarrow$ \\
    \cmidrule(r){3-5} 
           &  &  ET &  WT &  TC &  Per Slice \\
    \midrule
      \multirow{2}{*}{U-Net \cite{unet}}  &Baseline  & 77.04 & 90.27 & 82.53 & 17.07 \\
        &Ours   & 78.55 & 91.01 & \textbf{82.85} & \textbf{2.75}   \\
    \midrule
      \multirow{2}{*}{ TransBTS \cite{transbts}}   &Baseline  & 78.55 & \textbf{91.25} & 82.80  & 37.73  \\
        &Ours   & \textbf{79.01} & 91.03 & 82.18 & 3.55 \\    
    \bottomrule
  \end{tabular}}
  \caption{Ablation Study on Different Baselines.}
  \label{tab:ablation5}
    \end{subtable}

\vspace{-5pt}
    \caption{Ablation study of various aspects of our proposed Med-DANet V2 approach. }
    \vspace*{-0.2cm}
\label{ablation_study}
\vspace{-5pt}
\end{table*}

\section{Discussion and Conclusion}

In this paper, we present a study to explore the potential of dynamic inference in medical volumetric segmentation. 
We propose to unify the spatial dynamic adaption and dynamic model quantization to handle the slice heterogeneity in medical volume data.
From the perspective of data properties and model structure, we focus on the 3D MRI brain tumor segmentation and propose a novel framework named Med-DANet V2 with adaptive input selection and dynamic architectures to pursue the trade-off between segmentation accuracy and efficiency. 
In comparison with the previous work Med-DANet in this research direction, more fine-grained dynamic spatial adapting, and more flexible dynamic structural adapting are jointly incorporated to greatly promote the model efficiency.
Extensive experiments on two benchmark datasets for multimodal 3D MRI brain tumor segmentation demonstrate that our Med-DANet V2 reaches comparable or better performance than previous state-of-the-art methods with significantly less model complexity.

\noindent \textbf{Broader Impact and Limitation.} Our study offers a novel perspective and solution to realize efficient medical volumetric segmentation for clinical applications by holistically considering the unique characteristics of medical volume data along with adaptive model capacity adjustments. Our approach could be especially beneficial for time-sensitive medical diagnoses and treatment planning, where our framework's efficiency could lead to quicker and potentially more accurate clinical decisions. The proposed framework opens up a new avenue for achieving efficient volumetric segmentation in clinical settings, thereby serving as a catalyst for future research in this domain. 

However, one potential limitation could be that our current method for fine-grained, spatially-aware dynamic adaptation is restricted to isolating important foreground regions with regular geometries. This shortcoming identifies an avenue for future research, specifically the development of more nuanced dynamic input adaptation techniques capable of accurately identifying and segmenting foreground regions with irregular shapes.


{
\small
\bibliographystyle{plain}
\bibliography{reference}
}

\newpage
\appendix

\section*{Appendix}

\section{Visualization Results}

\subsection{Visual Comparison on BraTS 2019 Dataset}

\begin{figure*}[h]
\centering
\vspace{-12pt}
	\includegraphics[width=0.8\linewidth]{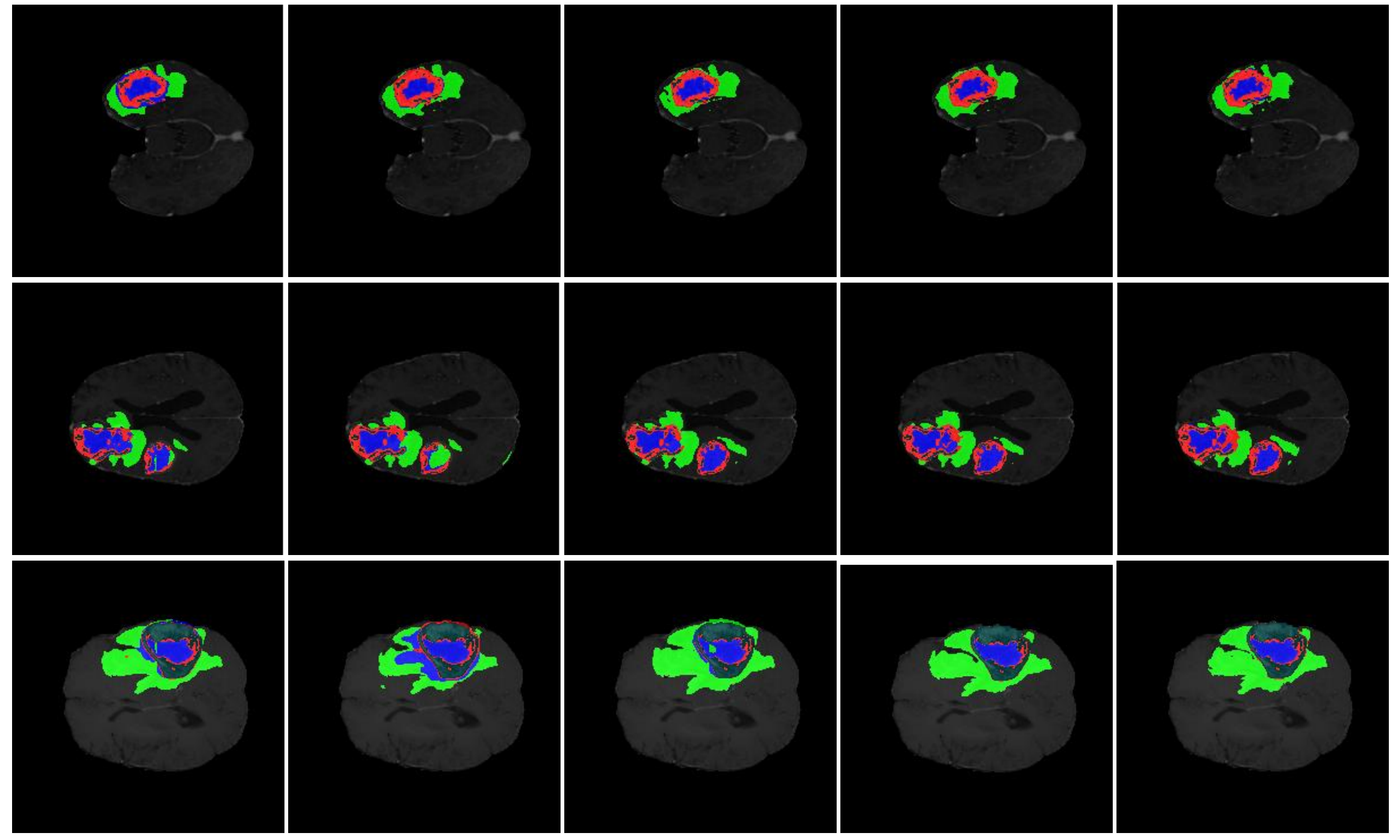}
     \begin{tabu} to 0.8\linewidth{X[1.0c] X[1.0c] X[1.0c] X[1.0c] X[1.0c]} 
        \scriptsize{3D U-Net} &  \scriptsize{VNet} &  \scriptsize{Att. U-Net} &  \scriptsize{\textbf{Ours}} &  \scriptsize{Ground Truth} \\
    \end{tabu}
	\caption{The Visual Comparison on BraTS 2019 dataset.}
	\label{fig_comparison}
\end{figure*}

The \textbf{qualitative analysis} on the segmentation performance of various methods are shown in Fig. \ref{fig_comparison}, including 3D U-Net, V-Net, Attention U-Net and the proposed Med-DANet V2. Due to the labels of the validation set are not available, we perform the five-fold cross-validation evaluation on the training set for a fair comparison. The visual results clearly demonstrate that our framework significantly enhances the delineation of brain tumors, yielding improved segmentation masks by focusing on the tumor regions that are more worth segmenting with the obtained optimal decision in terms of input resolution and quantization.

\subsection{Visualization of the Cropped Region}

\begin{figure*}[htbp]
\centering
	\includegraphics[width=0.6\linewidth]{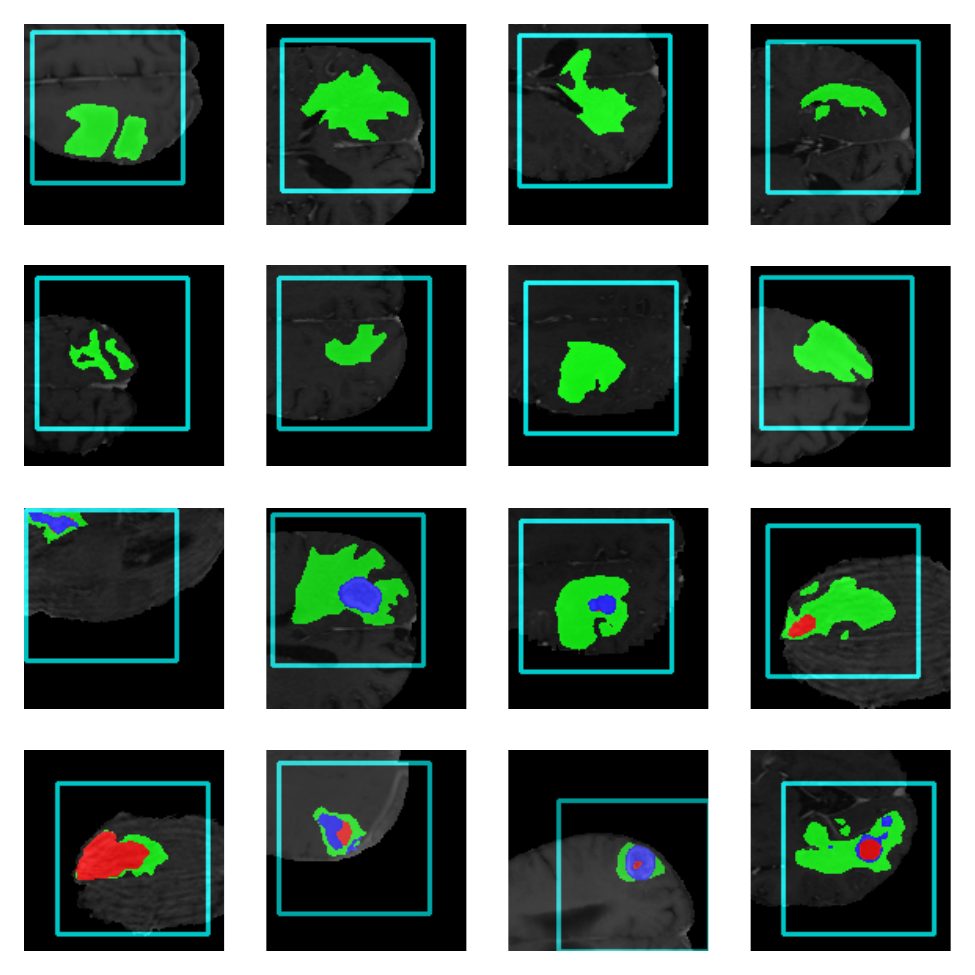}
	\caption{The Visualization of the Cropped Region.}
	\label{fig_crop}
\end{figure*}

The visualization of the cropped region on BraTS 2019 dataset is also shown in Fig. \ref{fig_crop}, where the cyan boxes indicate the cropped input determined by the Policy Network and Crop Position Network. 
It can be clearly observed that
our framework can adaptively select the specific cropping target among different slices based on the segmentation difficulty and more concentrate on the tumor-relevant regions for the cropped slices.

\end{document}